\documentclass[aps,floats,superscriptaddress,showpacs]{revtex4}
\usepackage{amsmath}

\usepackage{epsfig}
\usepackage{graphicx}
\usepackage{dcolumn}
\usepackage{bm}

\def\gapp{\lower.35em\hbox{$\stackrel{\textstyle>}{\sim}$}}
\def\lapp{\lower.35em\hbox{$\stackrel{\textstyle<}{\sim}$}}

\begin{document}
\bibliographystyle{apsrev}
%


\title{Dislocations in graphene}
\author{Ana Carpio}
\affiliation{ Departamento de Matem\'atica Aplicada, Universidad
Complutense de Madrid; 28040 Madrid, Spain}
\author{Luis L. Bonilla}
\affiliation{G. Mill\'an Institute for Fluid Dynamics, Nanoscience
and Industrial Mathematics, Universidad Carlos III de Madrid;
28911 Legan\'es, Spain, }
\affiliation{Unidad Asociada al Instituto de Ciencia de Materiales de Madrid, CSIC,}
\author{Fernando de Juan}
\author{Mar\'{\i}a A. H. Vozmediano}
\affiliation{Unidad Asociada ICMM-UC3M,
Instituto de Ciencia de Materiales de Madrid,\\
CSIC, Cantoblanco; 28049 Madrid, Spain.}

\date{\today}
\begin{abstract}
We study the stability and evolution of various elastic defects in
a flat graphene sheet and the electronic properties of the most
stable configurations. Two types of dislocations are found to be
stable: ``glide" dislocations consisting of heptagon-pentagon
pairs, and ``shuffle" dislocations, an octagon with a dangling
bond. Unlike the most studied case of carbon nanotubes, Stone
Wales defects are unstable in the planar graphene sheet. Similar
defects in which one of the pentagon-heptagon pairs is displaced vertically
with respect to the other one are found to be dynamically stable. Shuffle
dislocations will give rise to local magnetic moments that can
provide an alternative route to magnetism in graphene.

\end{abstract}
%
%
%
%
 \maketitle

\section{Introduction}

Graphene has become a very popular material since its recent
synthesis \cite{Netal05,ZTSK05} and characterization. Among the
most interesting properties related to the possible technological
applications are its high electron mobility
 and minimal conductivity at zero bias
\cite{GN07}.  Despite the high mobility of most of the graphene
samples, their mean free path of the order of microns \cite{Netal05}
implies the presence of defects. Very recent experiments performed on
suspended graphene \cite{BSetal08,DSetal08} indicate that, besides
the influence of the substrate, there must be intrinsic defects in
the samples.

The structure of disorder is also crucial to explain the magnetism
found in graphite samples \cite{Oetal07,Betal07}. It is now
clear that the intrinsic ferromagnetism is linked to defects in
the sample altering the coordination of the carbon atoms
(vacancies, edges or related defects)\cite{KM03}. One of the most
stable defects found in this work, shuffle dislocations, has an
unpaired electron that can contribute to the magnetic properties
of the sample.

Local disorder in graphene have been studied intensely and we
refer to the review article \cite{RMP08} for a fairly complete
list of references. A different type of disorder is provided by
the observation of ripples  in suspended graphene
\cite{Metal07,Metal07b} and in graphene grown on a substrate
\cite{Setal07,Ietal07}.

Inspired by the physics of nanotubes and fullerenes, curved
graphene has been modelled with curvature induced by topological
defects \cite{TT94,CR01,GGV01,CV07a,CV07b}. In these works it was
shown that conical singularities in the average flat graphene
sheet induce characteristic charge anisotropies that could be
related to recent observations \cite{Y08}.

Elastic and mechanical properties of graphitic structures  have
been studied intensely in the past, mostly in the context of
understanding the formation of fullerenes and nanotubes.  Very
little work has been done for the flat graphene sheet
\cite{FLK07,CK07} and topological defects have been often
excluded in these studies. In the fullerene literature it was established that the
formation of topological defects (substitution of a hexagonal ring
by other polygons) is the natural way in which the graphitic net
heals vacancies and other damages produced for instance by
irradiation \cite{LWetal05}. Among those, disclinations (isolated
pentagon or heptagon rings), dislocations (pentagon-heptagon
pairs) and Stone-Wales (SW) defects (special dislocation dipoles) were
found to have the least formation energy and activation barriers.
Dislocations and SW defects have been observed in carbon
structures \cite{Hetal04} and are known to have a strong influence
on the electronic properties of nanotubes. The possible role played by nanotube curvature 
so as to stabilize various defects is not yet clear.
Glide and shuffle dislocations in irradiated
graphitic structures have been described in \cite{EHB02}.
Experimental observations of dislocations have been reported very
recently in graphene grown on Ir in \cite{CNetal08}.

The purposes of this work are to discuss the formation and
stability of topological defects (mainly dislocations) in a flat
graphene sheet and to analyze the electronic properties of the
graphene samples in the presence of the most stable defects. This
paper addresses two aspects of physical reality -- elasticity and
electronics -- that  are often described  in very different
languages. We intend to reach a general audience and have included
brief pedagogical descriptions of the methods used in both
disciplines.

This paper is organized as follows: Section \ref{elastic}
explains the method used to study the formation and stability of
defects and it describes their stable configurations. We find two types
of stable dislocations, one with a dangling bond. Stone-Wales
defects are found to be unstable in the flat lattice whereas similar
defects in which one of the pentagon-heptagon pairs is displaced vertically
with respect to the other one are found to be dynamically stable. Section
\ref{electronic} gives a brief description of the tight binding
method and the physical information that can be extracted from it.
The electronic characteristics of the two dislocations are
derived. In section \ref{final} we present the conclusions and
future work.

\section{Periodized discrete elasticity and stability of defects}
\label{elastic}
\begin{figure}
\begin{center}
\includegraphics[width=8cm]{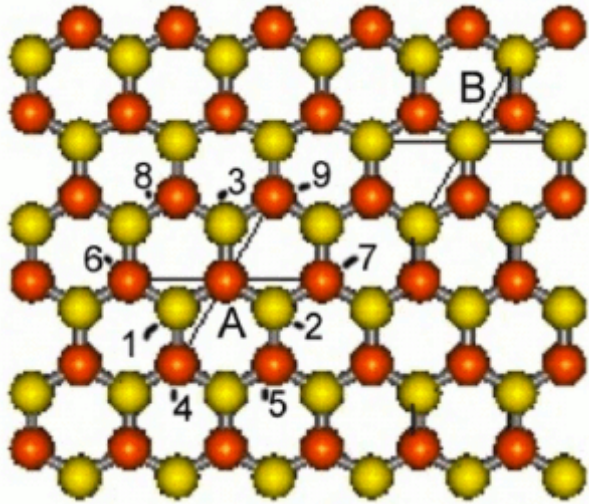}
\caption{(Color online) Neighbors of a given atom $A$. Only the
neighbors labelled 1, 2, 3, 4, 6, 7 and 9 are affected by the
difference operators $T$, $H$ and $D$ used in our discrete
elasticity model. } \label{figura6}
\end{center}
\end{figure}
In continuum mechanics, dislocations are usually described by the
equations of linear elasticity with singular sources whose
supports are the dislocation lines. To describe dislocations in 2D
graphene, we should have a more detailed theory which can be used
to regularize the corresponding point singularities. It is
possible to use ab initio theories as regularizers but, provided
dislocations are sparse and far from each other, there is a much
more economic and insightful alternative. We can discretize
appropriately linear elasticity on the hexagonal lattice and then
periodize the resulting linear lattice model to allow dislocation
gliding. The resulting model equations for the displacement vector
$(u'(n,t),v'(n,t))$ (written in primitive coordinates) are
\cite{CB08}:
\begin{eqnarray}
{\rho a^2\over 2} {\partial^2 u' \over \partial t^2}= {\lambda
+\mu\over  3}\, [(H-D)u' + (2H+D)v'] + T[(\lambda +3\mu)u'-
2(\lambda+\mu)v'], \label{e1} \\
{\rho a^2\over 2} {\partial^2 v'\over \partial t^2}=
{\lambda+\mu\over 3}\, [(H+2D)u' + (D-H)v'] + T[(\lambda +3\mu)v'-
2(\lambda+\mu)u']. \label{e2}
 \end{eqnarray}
where $n=(x,y)$ is a node $A$ or $B$ on one of the two sublattices
in Figure \ref{figura6}, $\rho$ is the mass density, $a$ is the
lattice constant and $\lambda$ and $\mu$ are the Lam\'e
coefficients which can be obtained from the elastic constants of
(isotropic) graphite in its basal plane, $C_{11}=C_{12}+ 2C_{66}=
1060$ GPa, $C_{12}=\lambda=180$ GPa, $C_{66}=\mu=440$
GPa.\cite{bla70} Note that $u'=(u-v/\sqrt{3})/a$ and $v' = 2v/(a
\sqrt{3})$ are nondimensional because the components of the
displacement vector in cartesian coordinates $(u,v)$ have units of
length. The difference operators $T$, $D$ and $H$ act on functions
of the coordinates $(x,y)$ of the node $A$ in Fig.~\ref{figura6}
according to the formulas:
\begin{eqnarray}
Tu'&=&g(u'(n_{1})-u'(A)) + g(u'(n_{2})-u'(A)) + g(u'(n_{3})-u'(A)),\label{T}\\
Hu' & = & g(u'(n_{6})-u'(A))+g(u'(n_{7})-u'(A)),  \label{H} \\
Du' &=& g(u'(n_{4})-u'(A)) + g(u'(n_{9})-u'(A)), \label{D}
 \end{eqnarray}
where $g$ is a periodic function, with period one, and such that
$g(x)\sim x$ as $x\to 0$. Note that the operator $T$ involves
finite differences with the three next neighbors of $A$ which
belong to sublattice 2, whereas $H$ and $D$ involve differences
between atoms belonging to the same sublattice along the primitive
directions $\mathbf{a}$ and $\mathbf{b}$, respectively. See Figure
\ref{figura6}. The same formulas hold if $(x,y)$ is an atom $B$ in
the other sublattice. Far from dislocation cores, the finite
differences are very small and close to the corresponding
differentials. If we Taylor expand these finite difference
combinations about $(x,y)$, insert the result in (\ref{e1}) and
(\ref{e2}) and write the displacement vector in cartesian
coordinates, we recover the equations of linear elasticity
\cite{CB08}.
\begin{figure}
\begin{center}
\includegraphics[height=5cm]{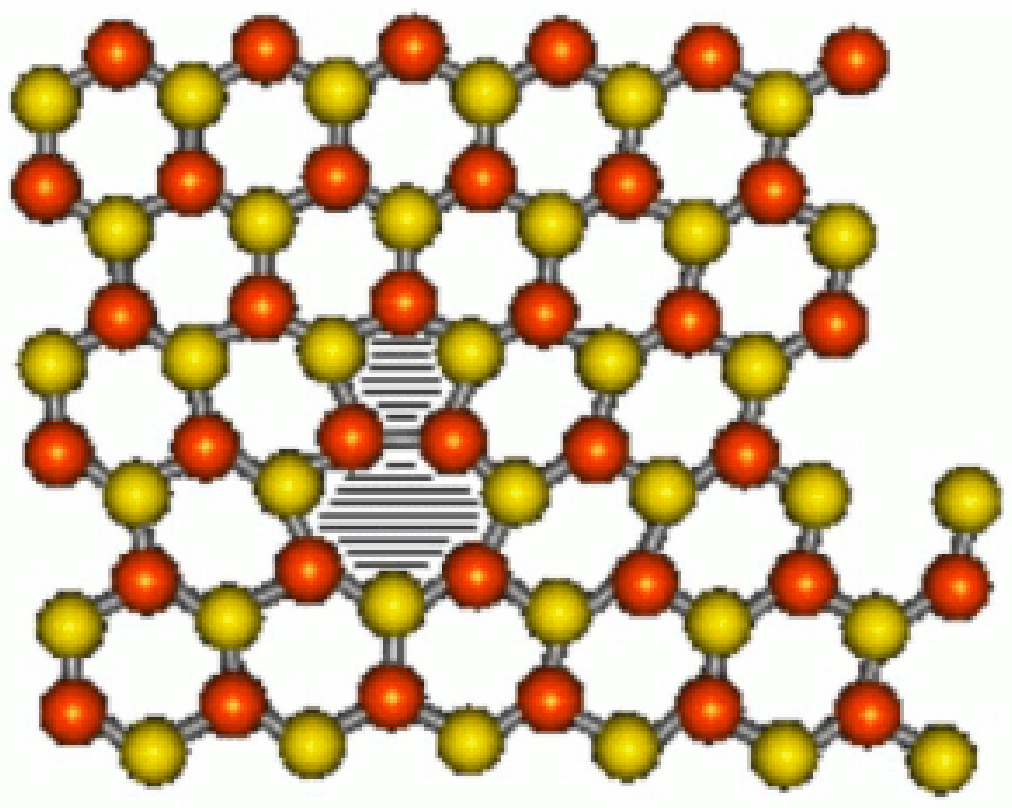}
\includegraphics[height=5cm]{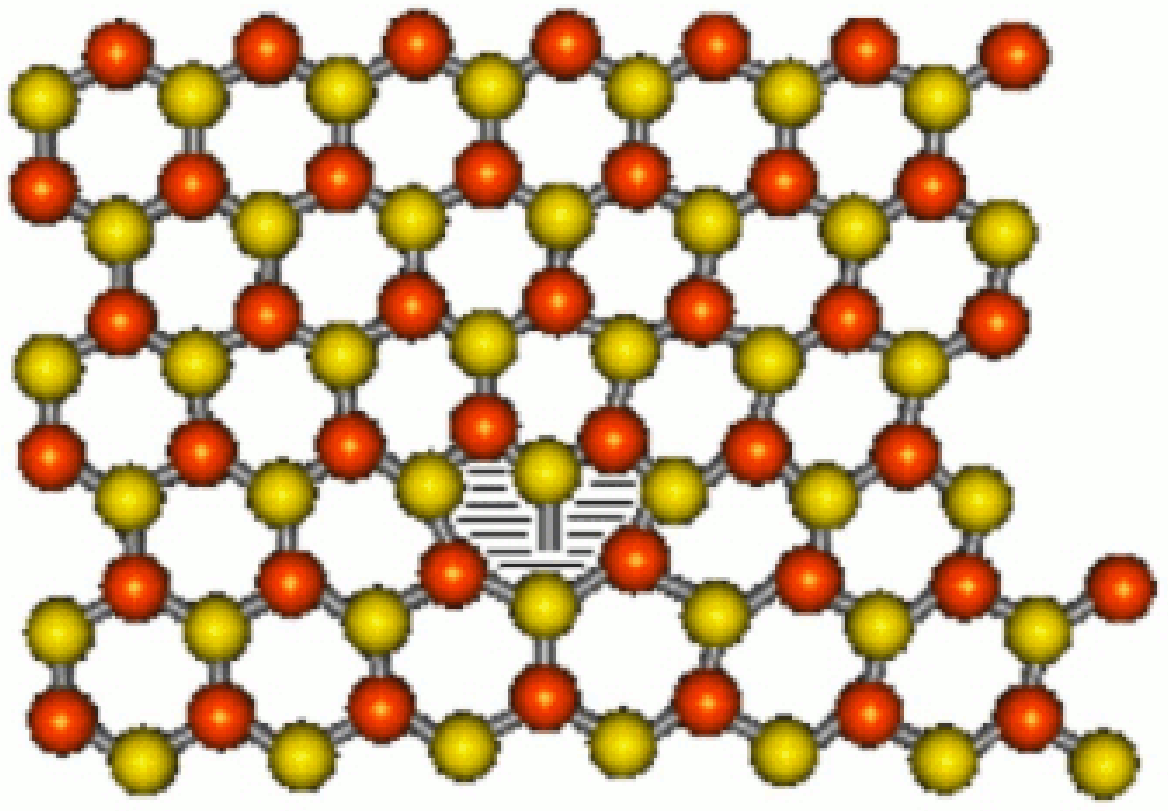}
\caption{ (Color online) Structure of the glide (left) and shuffle
(right) dislocations in the planar graphene lattice.}
    \label{ana}
\end{center}
\end{figure}
The role of the periodic function $g$ is to allow dislocation
gliding \cite{CB03,CB05,BCP07}. When a defect moves, a few atoms
change some of their nearest neighbors. We use the periodized
difference operators $T$, $D$ and $H$ in (\ref{e1}) - (\ref{e2})
instead of solving discrete elasticity with an updating algorithm
that keeps track of neighbor change. The equations of periodized
discrete elasticity (\ref{e1}) - (\ref{e2}) regularize linear
elasticity and allow for dislocation motion and for dislocation
nucleation \cite{PCB08}.

How do we find the defects in graphene that correspond to
different edge dislocations? We first substitute $(x,y)$ in the
elastic field of a dislocation (such as the edge dislocation of
page 57 of Ref.~\onlinecite{N67}) by $x=a\, (x'_{0}+l+y'_{0}/2 +
m/2)$, $y=a \sqrt{3}\, (y'_{0}+m)/2$. $l$ and $m$ are integer
numbers that allow the resulting displacement vector to be a
vector function of lattice points, which we denote by ${\bf U}
(l,m)$. The primitive coordinates $x'=x'_0 + l$, $y'=y'_0+m$ are
centered in an appropriate point $(x'_0,y'_0)$ which is different
from the origin to avoid the singularity in the elastic field to
coincide with a lattice point. We now solve an overdamped
periodized discrete elasticity model (in which second order time
derivatives are replaced by first order ones) with a boundary
condition given by ${\bf U}$ and with an initial condition also
given by ${\bf U}$. After a certain relaxation time, the solution
of the model evolves to a stable stationary configuration which
depends on the location of the origin $(x'_{0},y'_{0})$ and on
${\bf U}$. This stable configuration is also a stable
configuration of the original equations of the model (with
inertia).

By using the method just sketched, we have obtained that the same
dislocation solution of the equations of elasticity may have
different cores, which is a familiar fact in crystals with diamond
structure and covalent bonds, such as silicon; see page 376 in
Ref. \cite{HL82}. The stable configurations corresponding to one
edge dislocation are pentagon-heptagon defects (`glide'
dislocations) if the singularity is placed between two atoms that
form any non-vertical side of a given hexagon. If the singularity
is placed in any other location different from a lattice point,
the core of the singularity forms a `shuffle' dislocation: an
octagon having one atom with a dangling bond, as shown in
Fig.~\ref{ana}.

If we use the elastic field of an edge dislocation dipole as
initial and boundary condition, there are again different stable
configurations depending on how we place the dislocation cores. An
edge dislocation dipole is formed by two edge dislocations with
Burgers vectors in opposite directions. Let ${\bf E}(x,y)$ be the
displacement vector corresponding to the edge dislocation. If
${\bf U}={\bf E}(x-x_{0},y-y_{0}-l/2)-{\bf E}(x-x_{0},y-y_{0})$
($l=a/\sqrt{3}$ is the hexagon side in terms of the lattice
constant $a$), the stable stationary configuration is that of a
vacancy. If ${\bf U}={\bf E}(x-x_{0}, y-y_{0}-l)-{\bf
E}(x-x_{0},y-y_{0})$, a dynamically stable divacancy (formed by
one octagon and two adjacent pentagons) results. An initial
configuration corresponding to a Stone-Wales defect, ${\bf
E}(x-x_{0}-a,y-y_{0})-{\bf E}(x-x_{0},y-y_{0})$, is dynamically
unstable: at zero applied stress, the two component edge
dislocations glide towards each other and annihilate. If a shear
stress is applied in the glide direction of the two edge
dislocations comprising the SW defect, these defects either
continue destroying themselves or, for large enough applied
stress, are split in their two component heptagon-pentagon defects
that move in opposite directions \cite{CB08}.

Instead of a dislocation dipole, our initial configuration may be
a dislocation loop, in which two edge dislocations with opposite
Burgers vectors are displaced vertically by one hexagon side:
${\bf E}(x-x_{0}-a,y-y_{0})-{\bf E}(x-x_{0},y-y_{0}-l)$
($l=a/\sqrt{3}$ is the length of the hexagon side). In principle,
the dislocation loop could evolve to an inverse SW defect
(7-5-5-7). Instead, this initial configuration evolves towards a
single octagon. If we displace the edge dislocations vertically by
$l/2$, ${\bf E}(x-x_{0}-a,y-y_{0})-{\bf E}(x-x_{0},y-y_{0}-l/2)$,
the resulting dislocation loop evolves towards a single heptagon
defect \cite{CB08}.

\section{Electronic properties.}
\label{electronic}
 The electronic structure of the solids and most
of their low energy properties are dictated by the position of the
Fermi surface, its shape, and the amount of electrons available at
energies close to it. In the independent electron approximation,
valid when the kinetic energy of the electrons is much larger than
their mutual interactions, electronics is well described by
band theory. The latter gives two main outputs: geometry of the
Fermi surface and density of states at the Fermi level \cite{K96}.

The tight-binding approximation assumes that the electrons in the
crystal behave much like an assembly of constituent atoms. It
works by replacing the many-body Hamiltonian operator by a matrix
Hamiltonian. The solution to the time-independent single electron
Schr\"odinger equation is well approximated by a linear
combination of atomic orbitals. These form a minimal set of short
range basis functions $\phi_i$ -that we do not need to specify-
and the full wave function at site i is given by
$$\Psi_i=\sum_{ij}C_{ij}\phi_j.$$

The electron density at a lattice site $pq$ can be computed as
$$P_{pq}=2\sum_k^{occ}\sum_{pq}C_{pk}C^*_{qk}.$$
The tight binding energy is given by
$$E=\sum_{pq}P_{pq}h_{pq}=2\sum_k^{occ}\sum_{pq}C_{pk}C^*_{qk}h_{pq},$$
where $h{pq}$ is the element of the matrix Hamiltonian.

The advantage of the method is that matrix elements
$$h_{pq}=<p\vert H\vert q>=\int d{\bf r} \phi^*_p({\bf r})H\phi_q({\bf
r}),$$
$$S_{pq}=<\phi_p\vert\phi_q>=\int d{\bf r} \phi^*_p({\bf r})\phi_q({\bf
r}),$$ are not  explicitly calculated but approximated by
phenomenological parameters that depend on the geometry of the
lattice and the nature of the orbitals.
\begin{figure}
\begin{center}
\includegraphics[width=5cm]{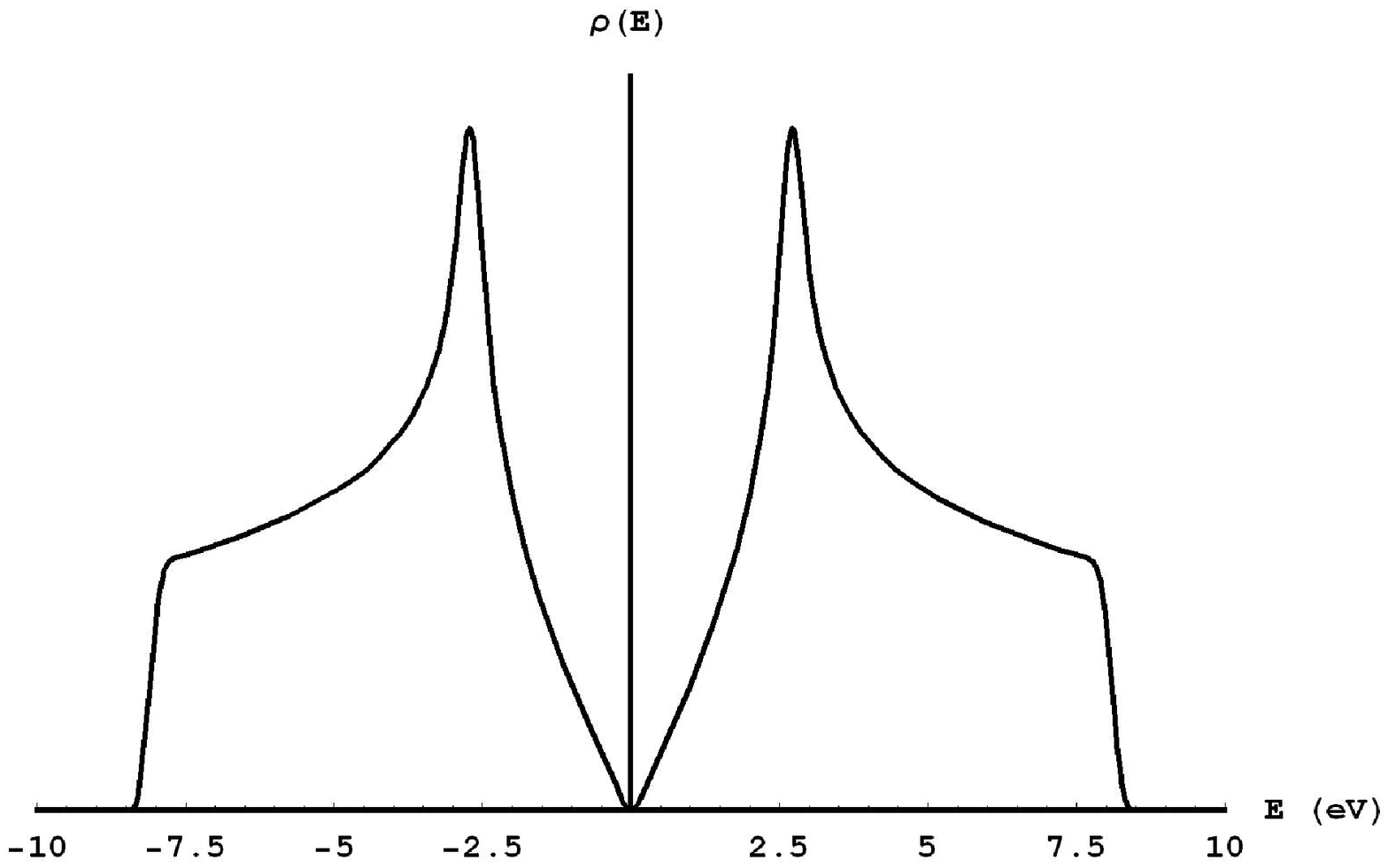}
\includegraphics[width=5cm]{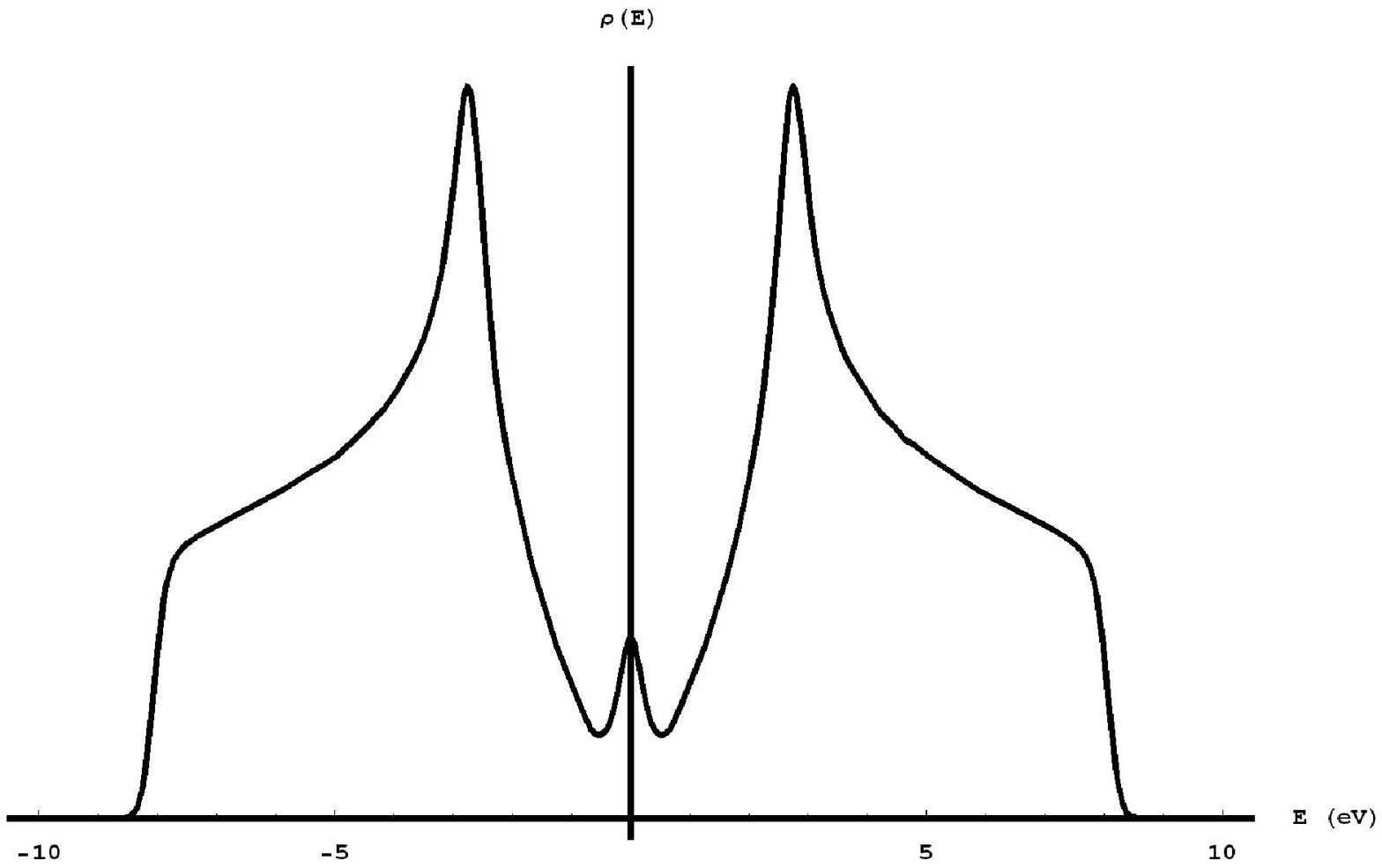}
\caption{Total density of states. Left Panel: Clean graphene. Right Panel:
Graphene ribbon with zigzag edges.} \label{0DOS}
\end{center}
\end{figure}
 The full strength of the tight binding approximation
is related with the perfect -discrete- translational invariance of
the periodic lattice.  The use of Bloch wave functions in Fourier
space allows a full description of the dispersion relation with
the only input of the overlapping integrals that can be indirectly
deduced from experiments. Since we are going to treat lattice
defects that break translational invariance we will stay in real
space and  adopt the simplest possible approximation: site
energies are set to zero and overlapping integrals are non-zero
only for nearest neighbor atoms. The hopping integral in graphene
is estimated to be of the order of $t \sim 2.7 eV$. In summary,
and in a very general sense, the electronic structure within the
tight binding approximation is obtained simply by defining a
lattice with links, and diagonalizing the Hamiltonian, a matrix
with elements $h_{ij}$ equal to $t$ if atom $i$ is linked to $j$
and zero otherwise. This is the calculation that we have
performed.

A full analysis of the tight binding structure of graphene can be
seen in the original paper \cite{W47} and in the reference book
\cite{SDD98}. Its main outcome is that the Fermi surface
reduces to two points and the density of states vanishes at the
Fermi energy which, in turn, determines the semimetallic character of the
material. The density of states is very important to characterize the electronic and
transport properties of the samples. Disorder can open a gap or,
more often, induce a finite density of states. Real samples have
localized states at (or about) zero energy which are induced
close to edges, vacancies, ad-atoms or other defects. These midgap
states can form very narrow bands where the electronic
interactions become important and may lead to electronic
instabilities, particularly ferromagnetism \cite{GKV08}.

The density of states of an ideal graphene sheet is shown 
in the left panel of Fig. \ref{0DOS}. It vanishes at the Fermi
energy what determines the semi-metallic character of the material.
Defects in the lattice very often induce states at zero energy. An
important class is that of edge states induced by certain boundaries
in finite lattices or real samples (graphene nanoribbons). Zigzag
(armchair) edges can be seen in the horizontal (vertical) borders
in Fig. \ref{figura6}. Zigzag edges with uncoordinated atoms
belonging to the same sublattice induce a number of zero energy edge
states proportional to the amount of unpaired lattice sites
\cite{Fetal96}. They are important in potential applications.
These energy states are localized at the edges as it can be seen in the
local DOS of Fig. \ref{ldos}. When studying electronic
properties via numerical simulations, it is important to
disentangle the low energy effects coming from the boundary from
those which are intrinsic to the defects under study. The density of
states of a graphene nanoribbon with zigzag edges is shown for
comparison in the right panel of Fig. \ref{0DOS}.

\vspace{0.5cm}

 {\it Electronic structure of single dislocations}

\vspace{0.5cm}

\begin{figure}
\begin{center}
\includegraphics[width=7.8cm]{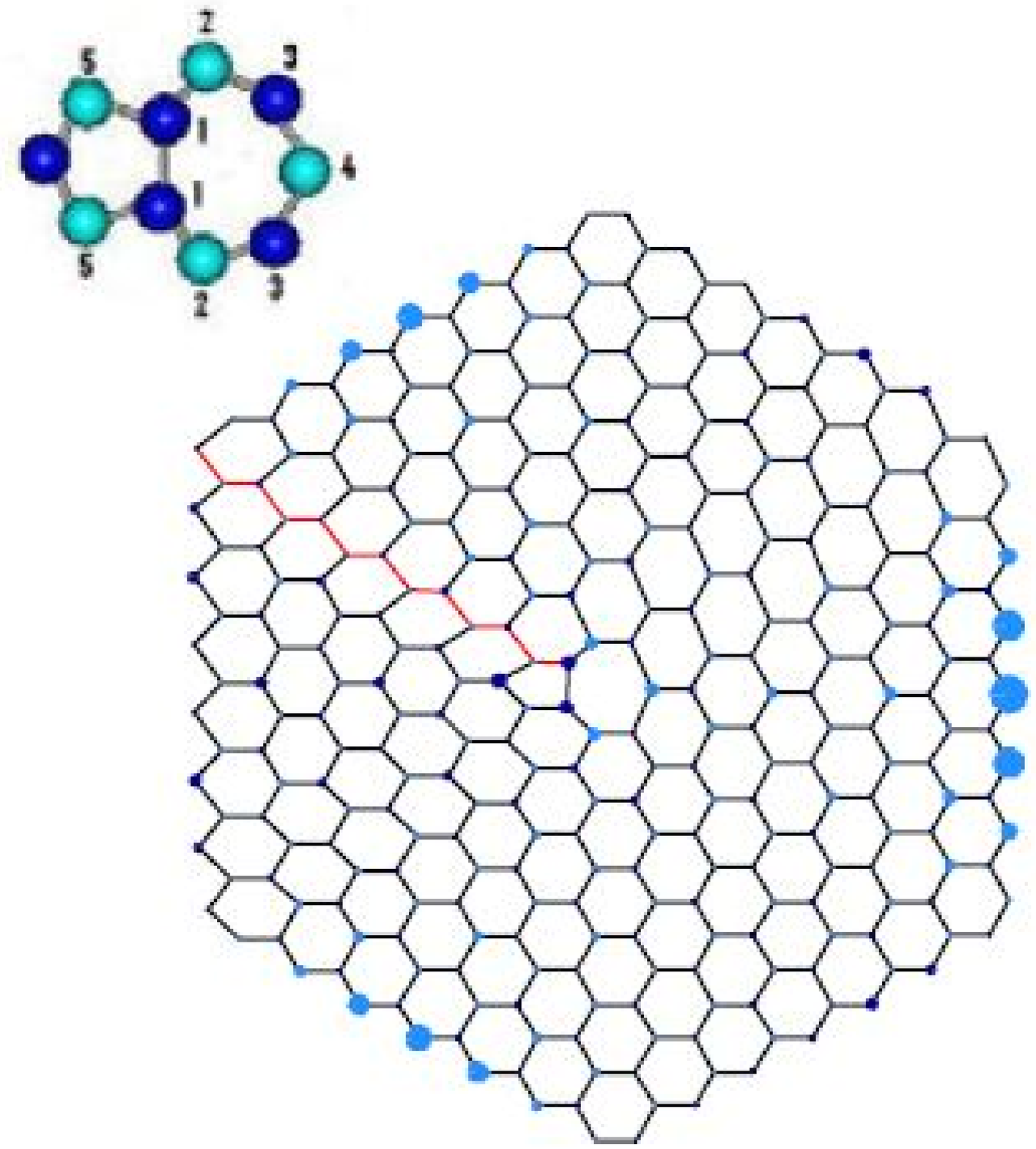}
\includegraphics[width=7.8cm]{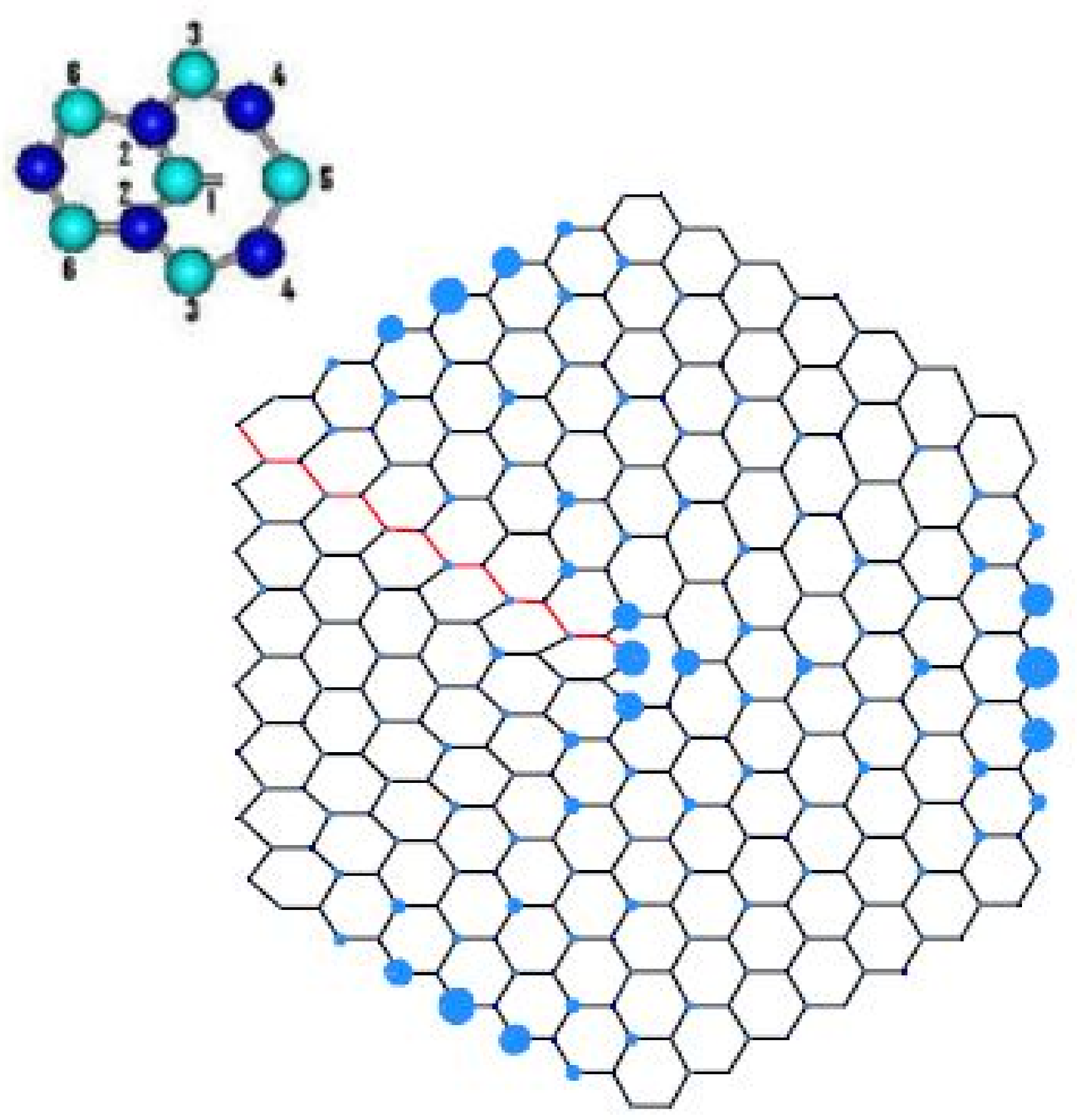}
\caption{(Color online) Left: Lattice structure and charge density
for a low energy eigenstate in the presence of a glide dislocation
shown in the inset.  Right: Same for the shuffle dislocation.}
    \label{lattice}
\end{center}
\end{figure}
%
\begin{figure}
\begin{center}
\includegraphics[width=7.8cm]{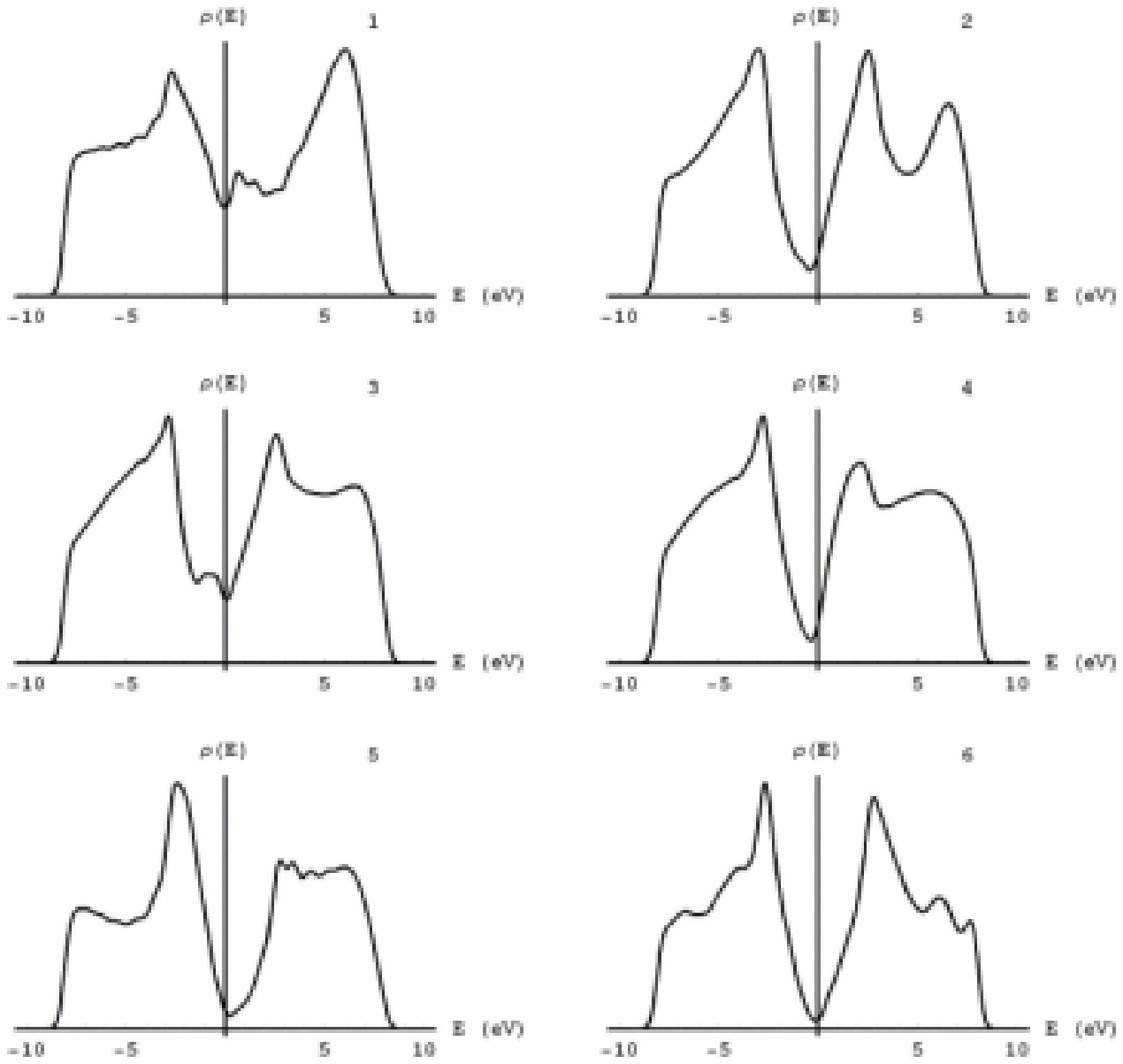}
\includegraphics[width=7.8cm]{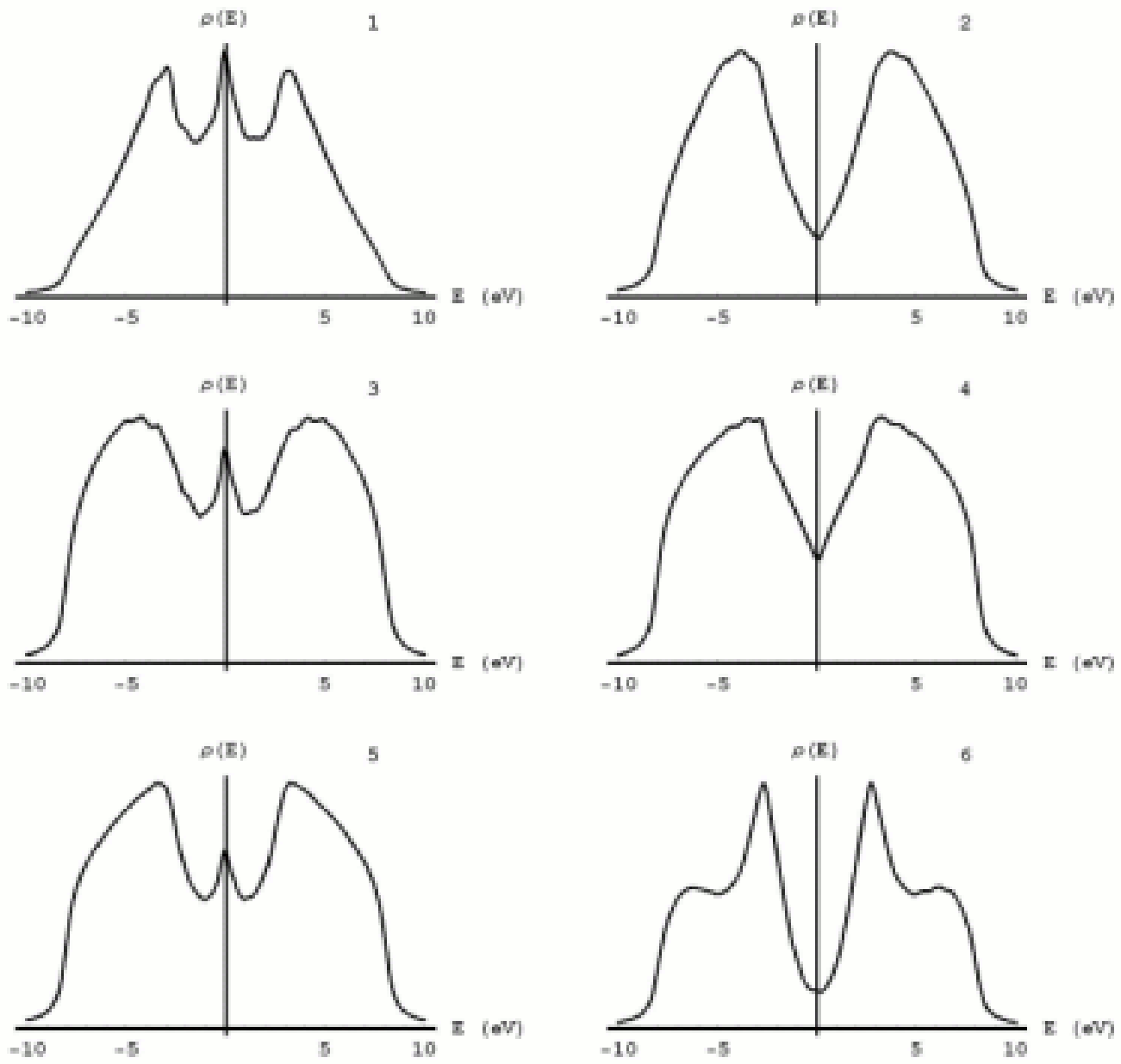}
\caption{Left: For the graphene sample with a
glide dislocation depicted in Fig. \ref{lattice}, local density of states 
at the numbered sites around the defect shown in the inset of Fig.
\ref{lattice}. Right: Same for the sample with a shuffle dislocation.}
    \label{ldos}
\end{center}
\end{figure}
As discussed in section \ref{elastic}, the ``glide" and ``shuffle" dislocations 
shown in Fig. \ref{ana} are stable in the graphene sheet. We have performed a tight 
binding calculation for these two types of dislocations. Fig. \ref{lattice}
shows the configuration of the lattice for the dislocations
depicted in the inset where the atoms that constitute the defect
are numbered. The extra rows of atoms characteristic of these edge dislocations
are shown in red. The area of the circles is proportional to the squared wave 
function for one of the lowest energy eigenvalues. The extra
charge appearing at the shuffle dislocation is due to  the
dangling bond attached to it.

In Fig.\ \ref{ldos}, we show the local density of states (LDOS) for
the five sites around the defect numbered in the inset of Fig.\
\ref{lattice} and for an extra site located at a certain distance
from the defect. The LDOS is drastically distorted at the defects
but rapidly recovers the normal shape away from the center of the
defect. The pentagon-heptagon pair (glide dislocation) breaks
the electron-hole symmetry of the lattice but the corresponding LDOS resembles
that of the perfect lattice shown in fig. \ref{0DOS}. The LDOS at
zero energy is not zero, but it has a minimum in all cases. The sixth
graph shows the LDOS at an atom located six lattice units apart
from the defect. This is the distance at which the influence of
the dislocation ceases to be noticeable.

The shuffle dislocation has a more pronounced effect on the LDOS.
As can be seen in Fig.\  \ref{ldos}, at zero energy there appear 
sharp peaks at the position of the dangling bond atom and at
neighboring sites of the same sublattice whereas dips in the LDOS
appear at the sites of the opposite sublattice. The distortion in
the LDOS decays faster with distance in the case of a shuffle
dislocation than in the case of a heptagon-pentagon pair. The right panel of
Fig. \ref{ldos} shows that the  density of states of the
perfect lattice is already recovered at position 6 of the inset
in Fig. \ref{lattice}, one lattice distance away from the defect.
The mid gap state induced by the defect is strongly peaked at the
defect position, similarly to what happens with the zigzag edges
states.  This type of dislocation does not break the electron hole
symmetry of the lattice.

\vspace{0.5cm}

{\it Defects of Stone Wales type}

\vspace{0.5cm}

\begin{figure}
\begin{center}
\includegraphics[height=6cm]{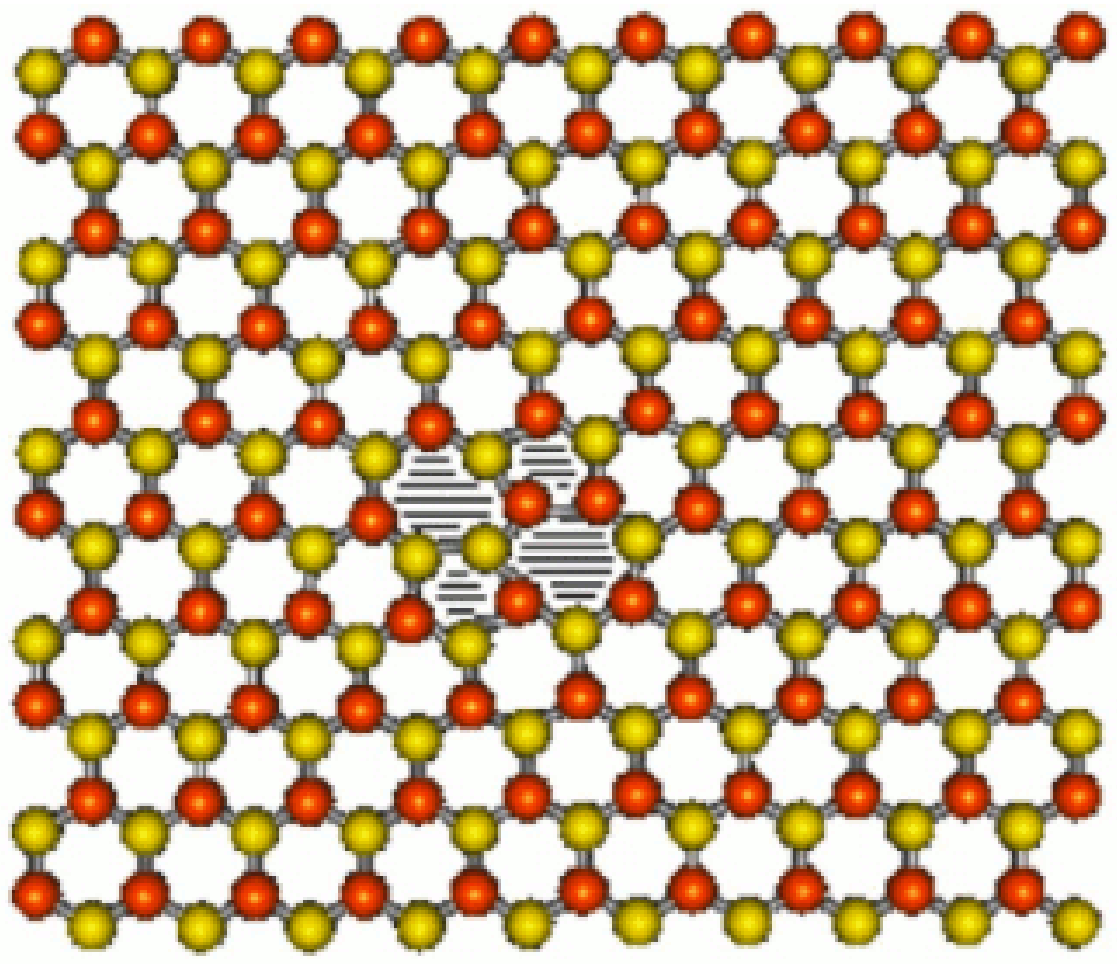}
\includegraphics[height=6cm]{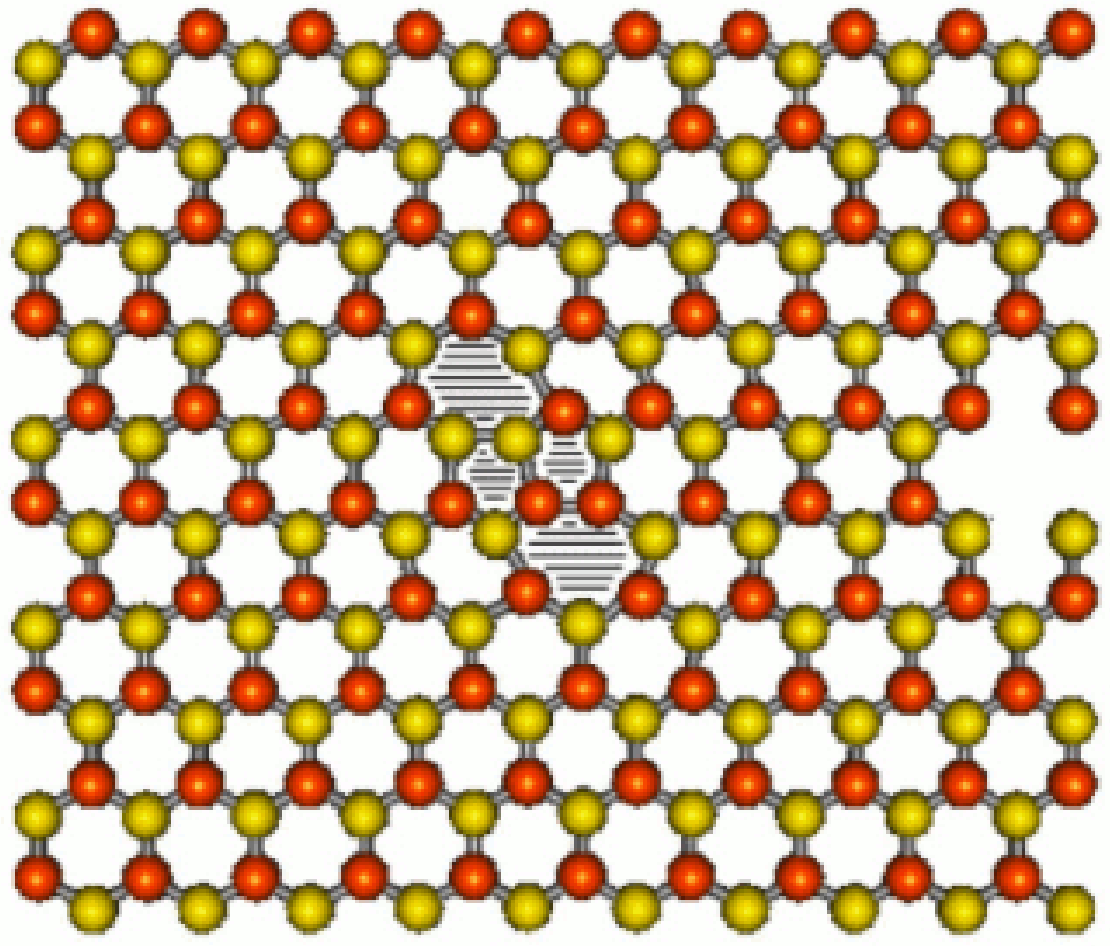}
\caption{(Color online) Left: Lattice structure of a Stone Wales
defect. Right: Same for the dislocation dipole described in the
text.}
    \label{SWAna}
\end{center}
\end{figure}
\begin{figure}
\begin{center}
\includegraphics[width=6cm]{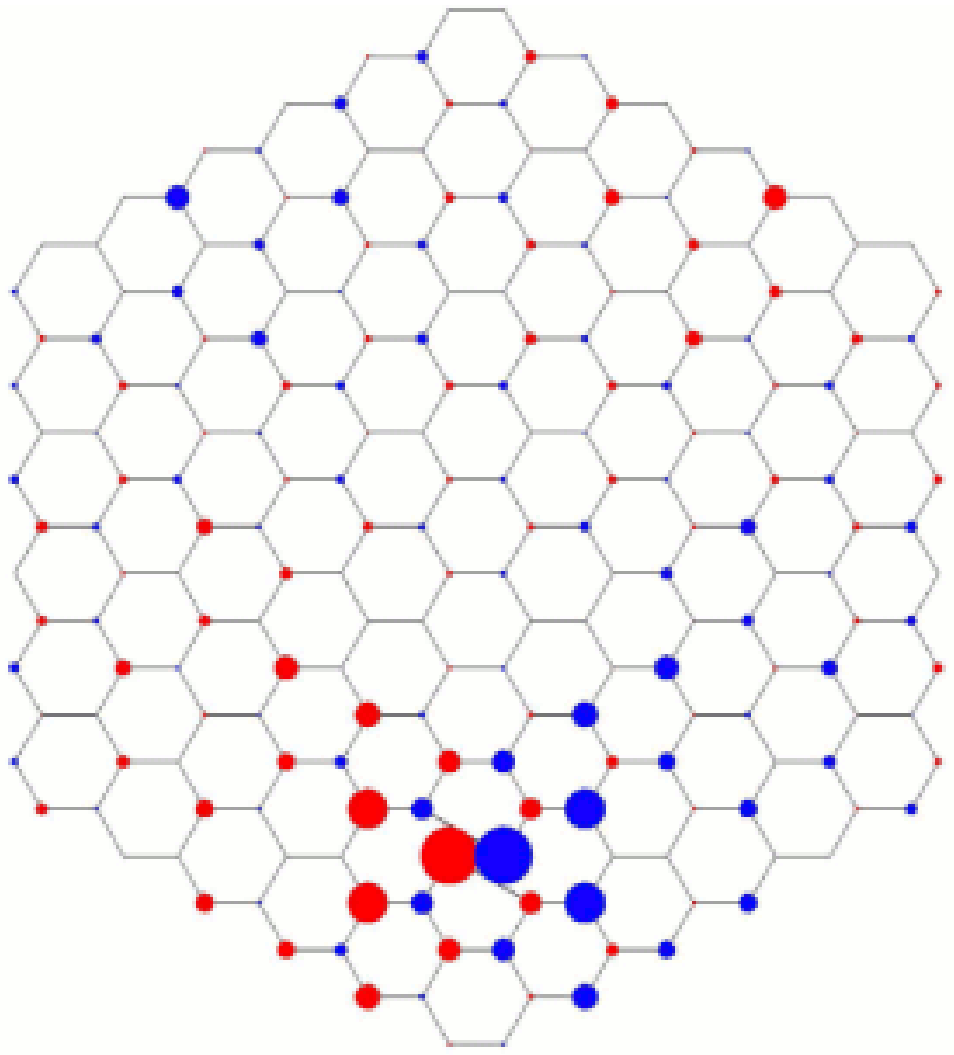}
\includegraphics[width=6cm]{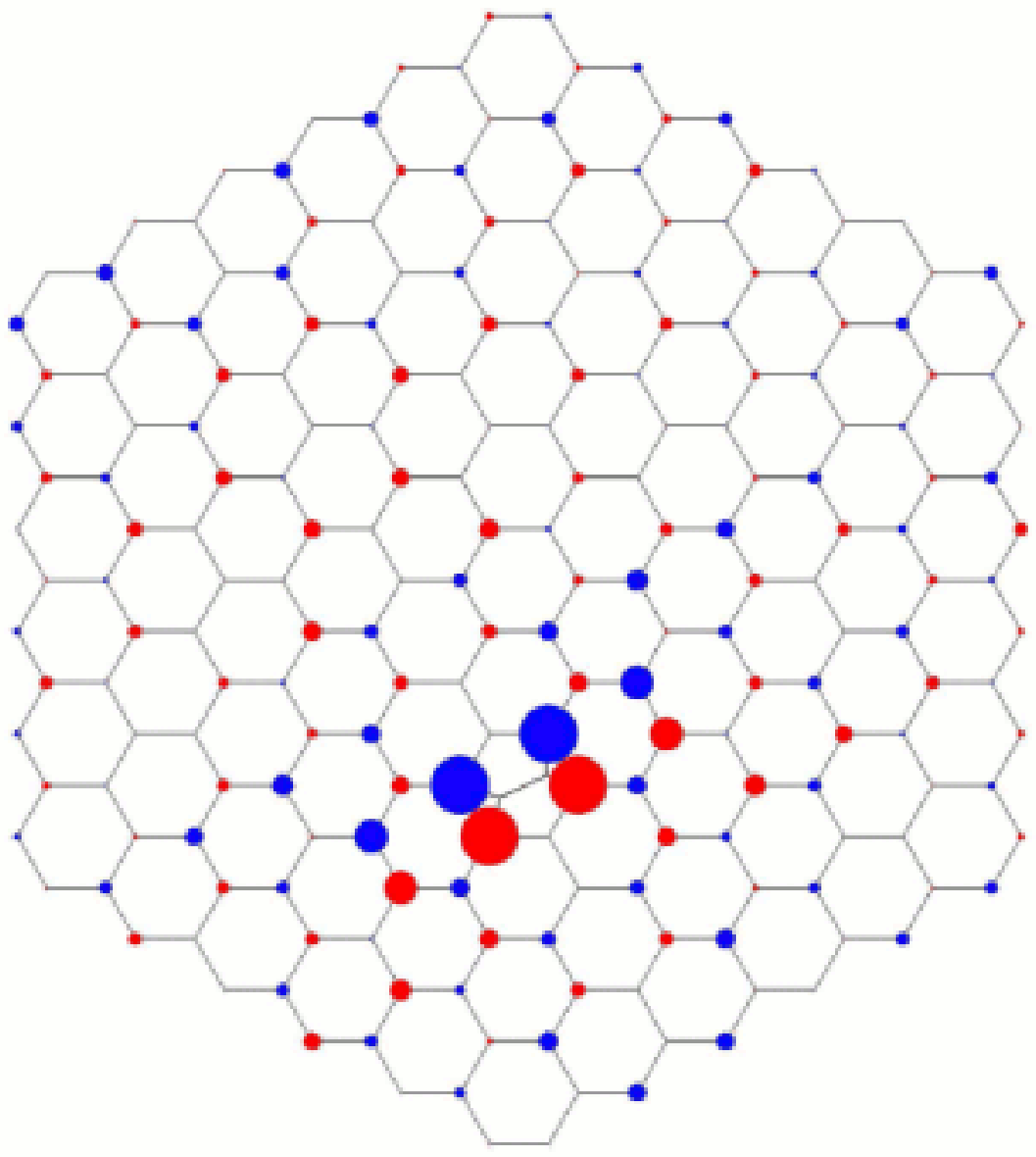}
\caption{(Color online) Left: Lattice structure and charge density
for the fourth positive-energy eigenstate in the presence of a
Stone Wales defect. Right: Same for the dislocation dipole
described in the text.}
    \label{SWlattice}
\end{center}
\end{figure}
\begin{figure}
\begin{center}
\includegraphics[width=4cm]{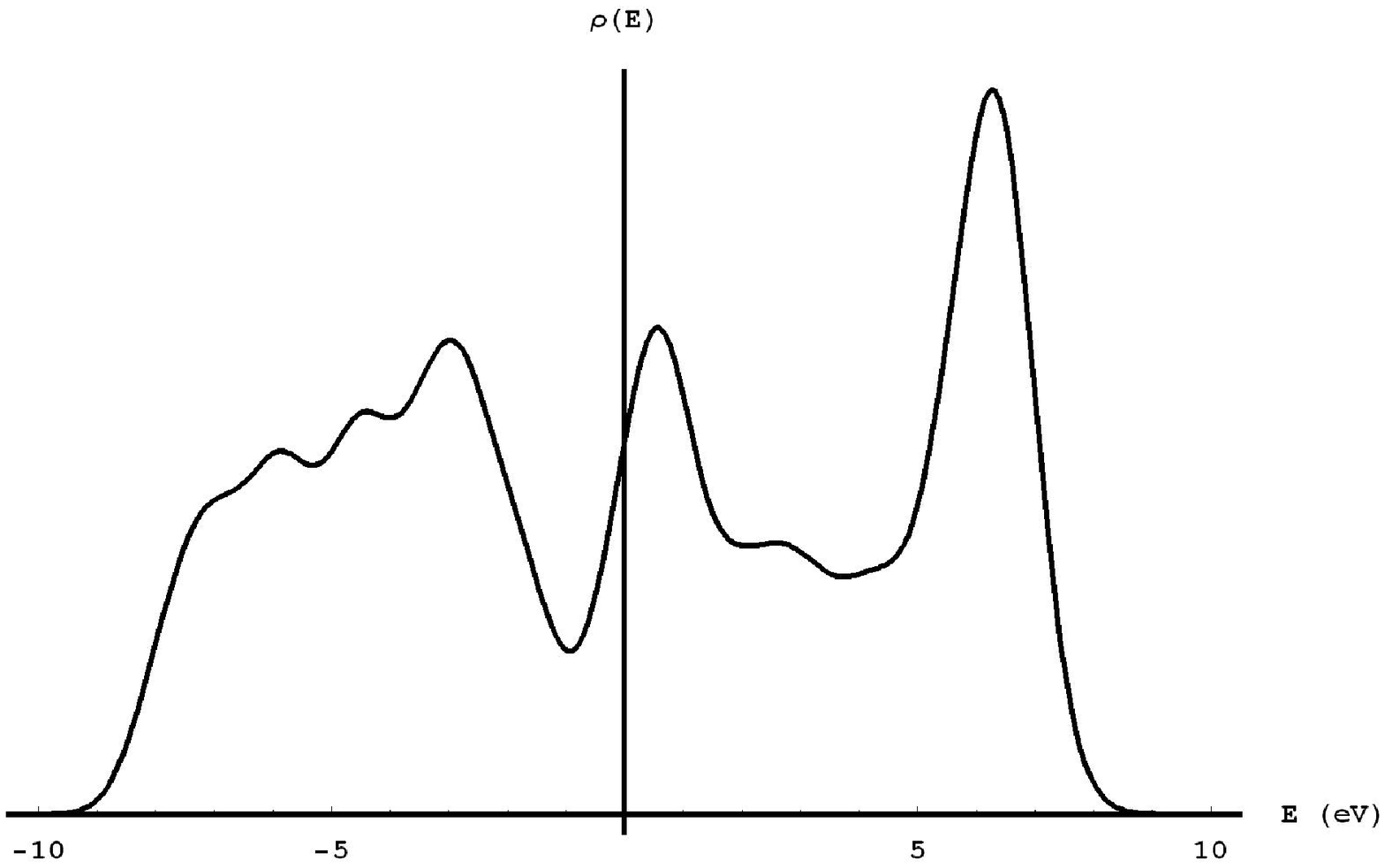}
\includegraphics[width=4cm]{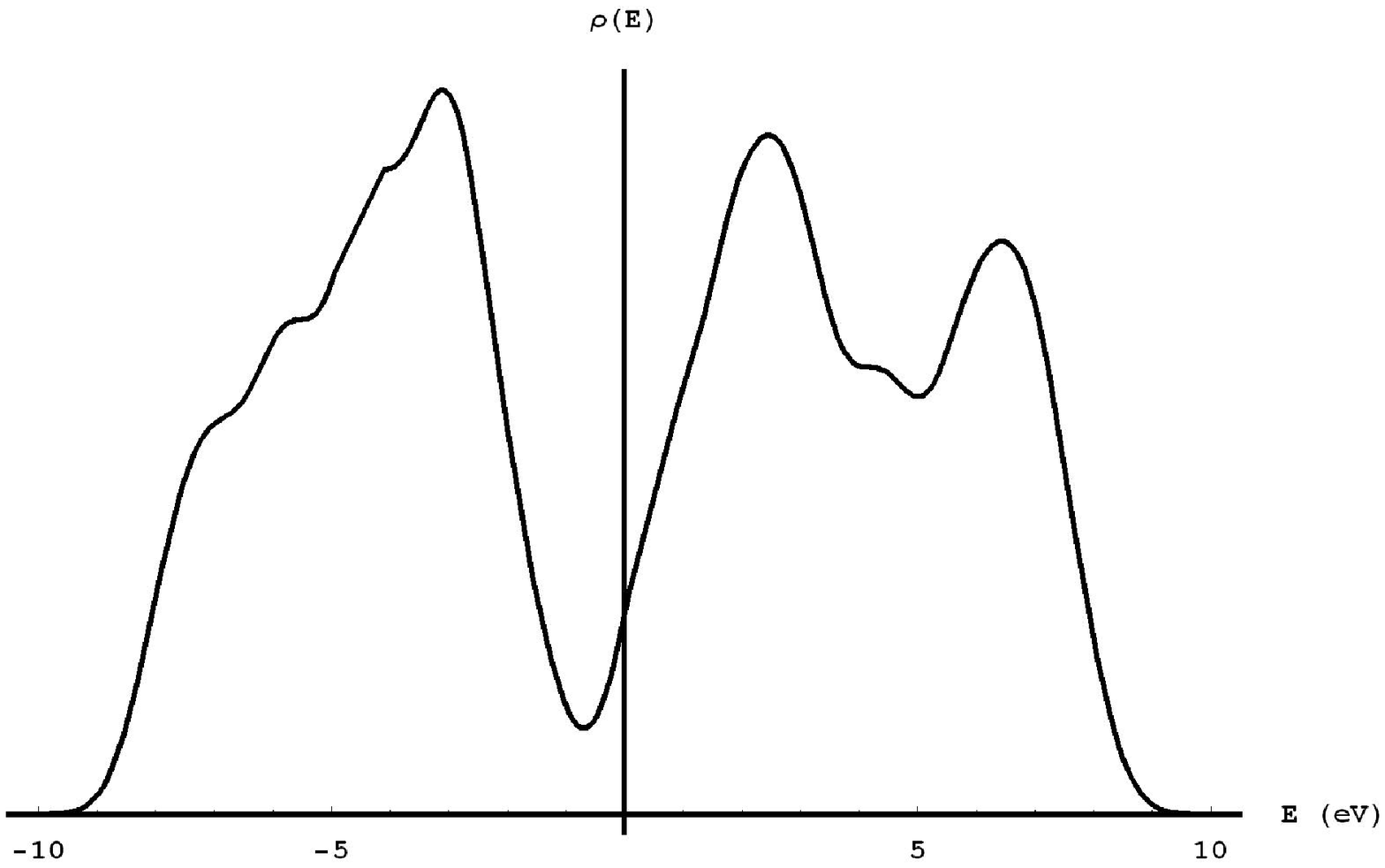}
\includegraphics[width=4cm]{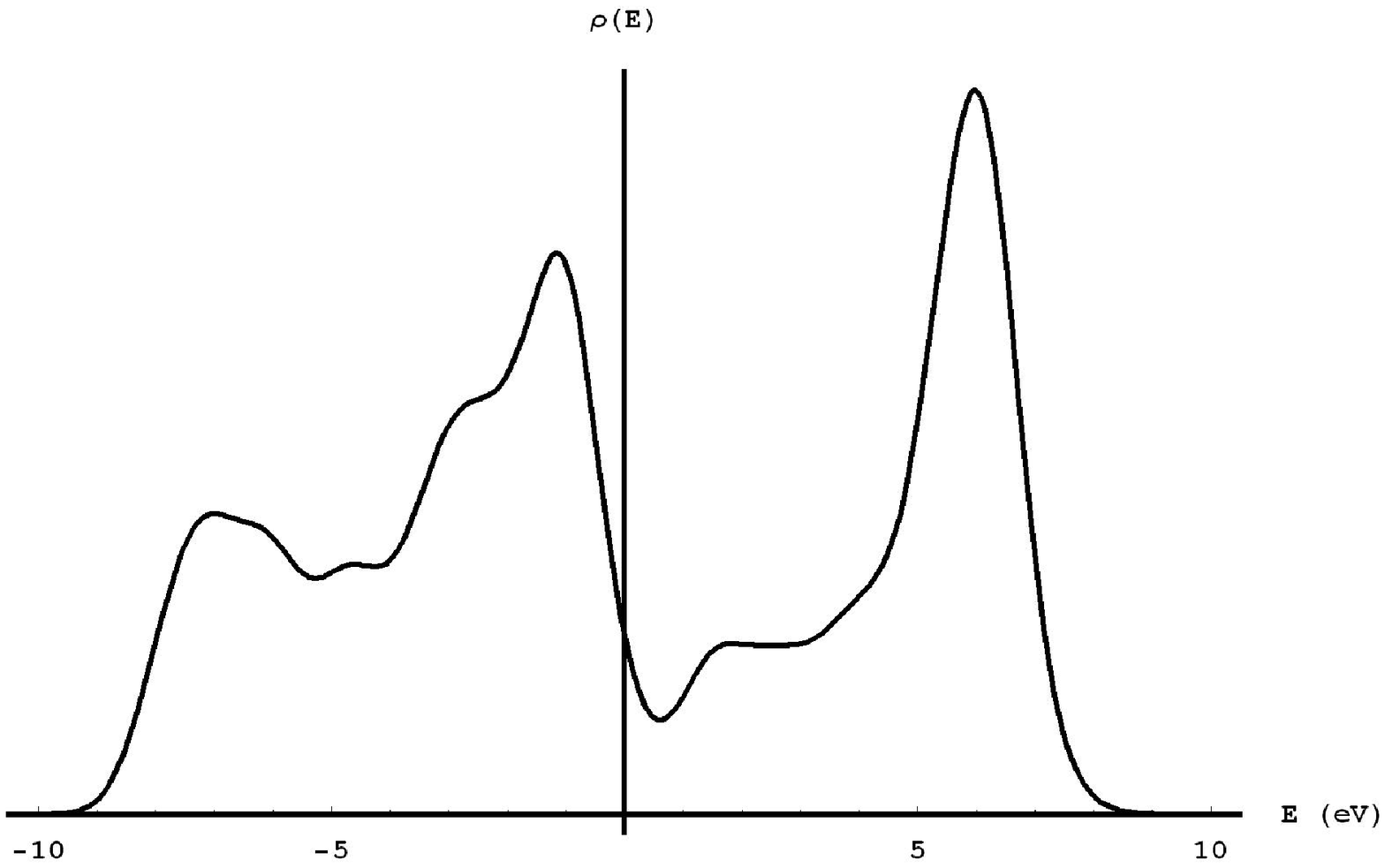}
\includegraphics[width=4cm]{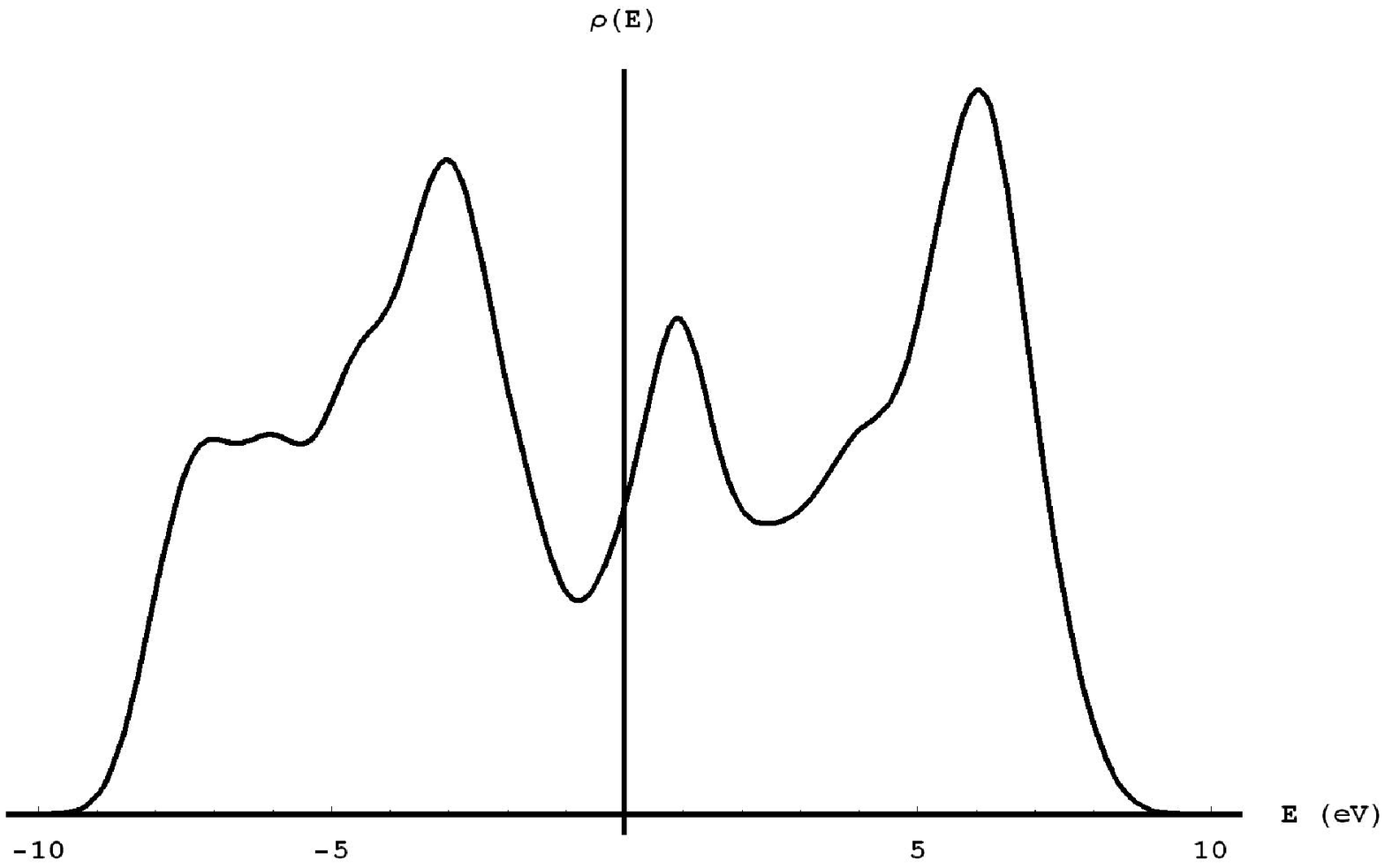}
\caption{ From left to right: Local density of states of a real SW
defect at the site shared by  the two adjacent pentagons and at its
nearest neighbor. LDOS of the dislocation dipole discussed in
the text at the site shared by the two heptagons and at its
nearest neighbor.}
    \label{SWldos}
\end{center}
\end{figure}

One of the best studied defects in the carbon nanotube literature
is the Stone Wales (SW) defect \cite{SW58}. It consists of two
heptagon-heptagon pairs that can be obtained by a ninety degree
rotation of a lattice bond. The resulting structure is shown in
the left hand side of  fig. \ref{SWAna}. These defects play a very
important role in the surface reconstruction  of irradiated
nanotubes \cite{ARC98} and affect their mechanical properties.
From the standpoint of elasticity, they can be seen as two
identical edge dislocations that have opposite Burgers vectors and
share the same glide line. They have been found to be dynamically
unstable: their component edge dislocations glide towards each other and annihilate, 
leaving the undistorted lattice as the final configuration
\cite{CB08}. A  type of defect whose final configuration is very
similar -- two heptagon-pentagon pairs -- is shown in the right panel
of Fig. \ref{SWAna}. It is a dislocation dipole whose two edge dislocations 
with opposite Burgers vectors are displaced vertically by one lattice unit. 
By solving the periodized discrete elasticity model of Section \ref{elastic}, we 
can show that this configuration is dynamically stable. The electronic structure of these 
two defects is depicted in Figs.\ \ref{SWlattice} and \ref{SWldos}. These dipole
defects induce a stronger local distortion of the charge density than single
dislocations. While the real SW defect does not alter the structure
of the lattice edges, the other dislocation dipole has two extra atoms as compared to the 
perfect lattice and therefore it alters the structure of its edges. This is clearly visible in 
Fig.~\ref{SWAna}. The presence of these defects can affect the electronic 
properties of real samples.

\section{Conclusions and discussion}
\label{final}

We have used a regularization of the continuum elasticity on
the honeycomb lattice to explore the stability and evolution of
topological defects. Two types of dislocations are stable:
pentagon-heptagon pairs (`glide' dislocations) and `shuffle'
dislocation: an octagon having one atom with a dangling bond. They
are shown in  Fig.~\ref{ana}. Both defects induce distortions in
the local density of states at low energies that decay rapidly
with the distance to the defect. The presence of a dangling bond
in the shuffle dislocations drastically enhances these effects but,
as in the case of zigzag states, the low energy states are very localized.

The  main physical effect of the shuffle dislocations will be
related with the nucleation of magnetic moments at the dangling
bonds. Work in this direction is in progress.

Regarding configurations of edge dislocation dipoles in discrete elasticity, 
vacancies and di-vacancies are stable but Stone-Wales defects are 
dynamically unstable. This situation is to be confronted with what happens in
the carbon nanotubes where SW defects are stable. This points to
the idea that curvature and geometry play a role in their stabilization. We
are also working in this direction.

A defect similar to the SW consists of a dislocation dipole whose 
component dislocations are displaced one lattice unit. This defect
is dynamically stable and can give rise
to a large local distortion of the electronic density. The
defects discussed in this work are very likely to be present in
real samples of both graphene and nanoribbons. They will affect
the transport properties of the samples and they will also alter the
configuration of the sample edges. This must be taken into
consideration in the cases when perfect tayloring of the edges is
important.

\acknowledgments This research was supported by  the Spanish MECD
grants MAT2005-05730-C02-01, MAT2005-05730-C02-02,
FIS2005-05478-C02-01 and by the Autonomous Region of Madrid under
grants S-0505/ENE/0229 (COMLIMAMS) and CM-910143 and by
PR27/05-13939. The {\it Ferrocarbon} project from the European
Union under Contract 12881 (NEST) is also acknowledged.

\bibliography{dislocations_r}

\end{document}